\documentclass[]{elsart}

\usepackage{natbib, graphicx}

\usepackage{amssymb}
\journal{Journal of Informetrics}
\begin{document}
\begin{frontmatter}

\title{Word statistics in Blogs and RSS feeds: Towards empirical universal evidence}

\author[lambi]{R. Lambiotte},
\ead{Renaud.Lambiotte@ulg.ac.be}
\author[lambi2]{M. Ausloos} and
\author[mike]{M. Thelwall}

\address[lambi]{CREEN, Universit\'e de Li\`ege, B5 Sart-Tilman, B-4000 Li\`ege, Belgium}
\address[lambi2]{GRAPES, B5 Sart-Tilman, B-4000 Li\`ege, Euroland}
\address[mike]{SCIT, University of Wolverhampton, Wulfruna Street, Wolverhampton WV1 1SB, UK}

\begin{abstract}
We focus on the statistics of word occurrences and of the waiting times  between such occurrences in Blogs. Due to the heterogeneity of words' frequencies, the empirical analysis is performed by studying classes of "frequently-equivalent" words, i.e. by grouping words depending on their frequencies. Two limiting cases are considered: the dilute limit, i.e. for those words that are used less than once a day, and the dense limit for frequent words. In both cases, extreme events occur more frequently than expected from the Poisson hypothesis.
These deviations from Poisson statistics reveal non-trivial time correlations between events that are associated with bursts of activities. The distribution of waiting times is shown to behave like a stretched exponential and to have the same shape for different sets of words sharing a common frequency, thereby revealing universal features.

\end{abstract}

\end{frontmatter}

\section{Introduction}

Web logs, also known as Blogs, have become an influential medium \cite{ham,glance,thelwall}, that encompasses a  broad variety of subjects, e.g. politics and science, and are participative by nature. They involve a huge number of interacting users that belong to several layers of the population, from topic specialists to average people. This variety suggests that Blogs could be an efficient information source for identifying, tracking and modeling the spread of ideas and opinion formation, for example in public debates over political questions. Indeed, the democratic nature of Blogs allows us to examine how trends develop from the interactions of decentralized bloggers and to follow dynamic opinion changes over a wide and diverse sample of the population. This is in contrast with the main media where relatively few journalists are involved. 
Precise knowledge of word statistics in Blogs is consequently of interest in order to make coherent statistical tests for automatically detecting critical events, e.g.,  trends or media shocks \cite{kleinberg1,kleinberg2}. 

The most basic time statistics ignoring correlations between events can be modeled by Poisson distributions. This distribution concerns independent events: the number $n$ of events arriving during some time interval $\Delta$ occurs with a probability
\begin{equation}
\label{0}
 P(n|a)=\frac{a^n}{n!} e^{-a},
 \end{equation}
where  $a$ is the arithmetic average number of events during this time interval. Moreover, the distribution of waiting times between 
two successive Poisson events is the  negative exponential: 
\begin{equation}
\label{poisson}
f(\tau)= \tau_{c}^{-1} \exp{(- \tau/ \tau_{c})},
\end{equation}
where $\tau_c = \Delta / a$ is the average characteristic waiting time between events. This distribution is well-known to apply to nuclear disintegration but it has also been used for describing the time gaps between shoppers entering a store \cite{shop}, the number of failure of products 
 \cite{gregory}, the number of terrorist acts \cite{irak} as well as the number of airplane accidents as a function of time \cite{lambi}. 
 An  increasing amount of empirical evidence indicates, though, that human activity patterns do not fit this model. It has been shown by many other authors that human processes are rather heterogeneously distributed in time, with short periods of high activity \cite{kleinberg1,kleinberg2,willinger}, or bursts, separated by long periods of inactivity \cite{bar,bara,dewes,paxson,deszo,vazquez,ebeling,gop,sab}. This heterogeneity is characterized by a distribution of waiting times which deviates from the exponential (\ref{poisson}) and which, usually, presents a  so-called heavy tail. 
 
 In this paper, we focus on the statistics of such waiting times between word occurrences in Blogs (and other similar periodically updated web sources) and also on the statistics of the number of word occurrences per day. To do so, we focus on texts published in 68022 RSS feeds during a period of 214 days and analyze two limiting cases. On one hand, we focus on very rare "events", namely words that occur on average less than once per day.  It is shown that the frequency of words is very heterogeneous, so that the time statistics have to be measured in classes of "frequently-equivalent" words, i.e. words are discriminated through their total number of occurrences during the whole time period. This discrimination allows us to show that the distribution of waiting times deviates from the exponential (\ref{poisson}), i.e. it is fitted by a stretched exponential and therefore presents an overpopulated tail.  The deviation from the pure exponential is evaluated with the quantity $\zeta$ that measures the importance of the second moment of the time statistics. Interestingly, it is found that the shape of the distribution as well as the value of $\zeta$ do not depend on the class of words in which they are measured.  On the other hand, we focus on events that occur many times per day on average. In that case, scaling laws are applied in order to smoothen the empirical results. Deviations from the Poisson statistics (\ref{0}) are also found. Consequently, our results not only confirm that the dynamics of topics in Blogs present bursts of activity  \cite{kleinberg1,kleinberg2,willinger} but they also provide tools in order to measure the importance of such bursts by comparing the empirical word statistics to a Poisson uncorrelated process.

\section{Data description}

\subsection{RSS format}
Really Simple Syndication (RSS) is an XML  application designed to deliver brief summaries of the most recent updates of web sites \cite{ham}, although it is flexible enough to incorporate other applications such as reporting updates in digital libraries or search engine databases. Users with RSS reader software can subscribe to a range of RSS feeds based upon their interests, perhaps including favourite Blogs, some news sites or some special interest sites. The RSS reader  will typically check each feed hourly and report to the user whenever new content is found. Each RSS feed contains a list of the most recent site updates, stored as separate XML ÔitemsÕ. When new content is added to the site, a new item will be added to the feed and the oldest one removed. Hence, when checking for updates, the RSS reader needs to parse feeds for items and report only items that are new, i.e. which were not in the feeds when they were previously checked.

RSS is also an attractive format for large scale data collection and analysis because it is typically concise and easy to parse text. In addition, its contents are easily time stamped so that time series can be generated. In contrast, web pages are typically much less concise and much harder to parse. Moreover, time series are difficult to generate from such web pages because they typically reveal at best a Ôlast modifiedÕ date (that is not automatically updated by the author). Like RSS, Blogs are more amenable to time series generation because each posting is dated and old postings are not normally modified. 

\begin{figure}
\includegraphics[angle=-90,width=4.5in]{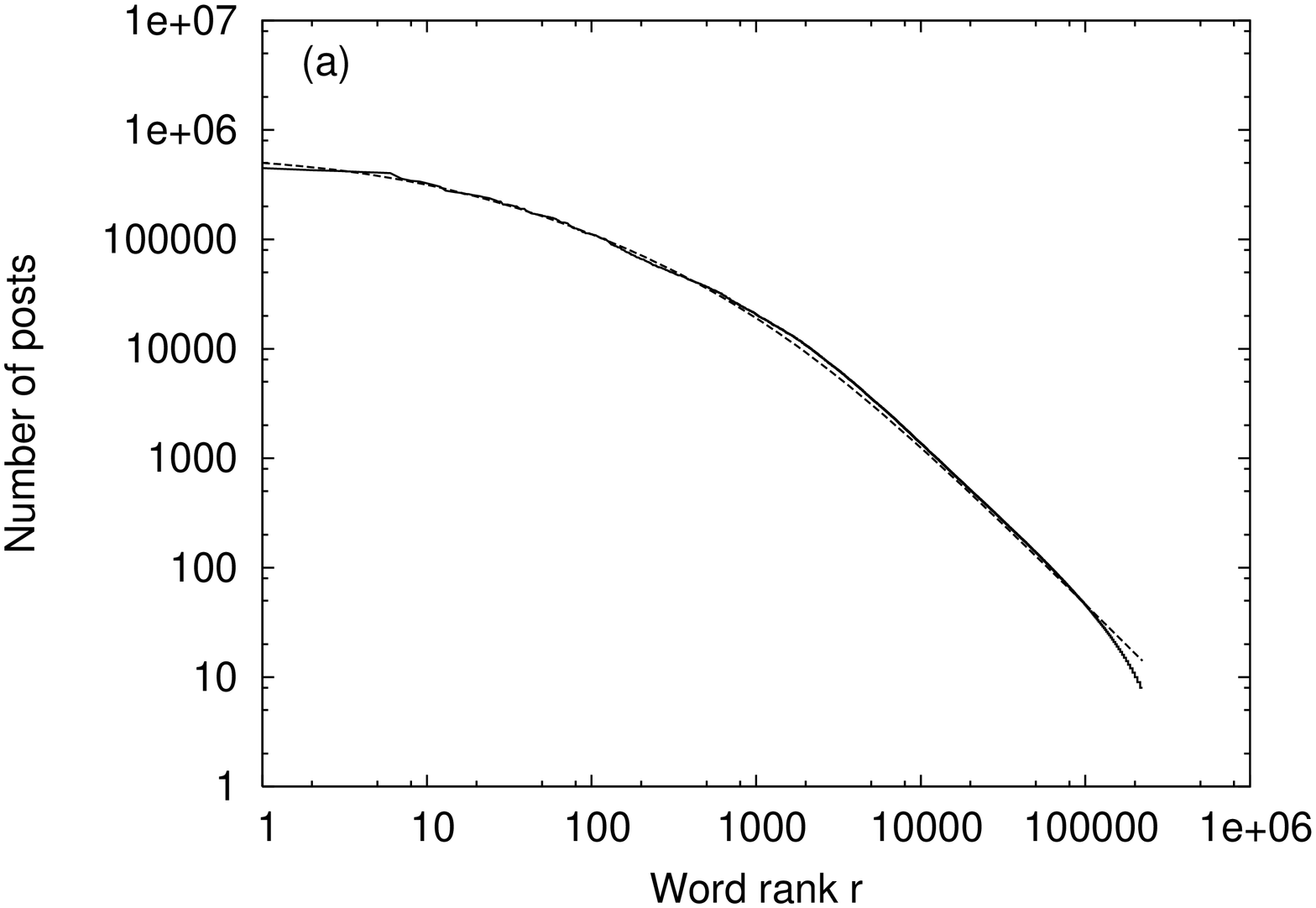}

\includegraphics[angle=-90,width=4.5in]{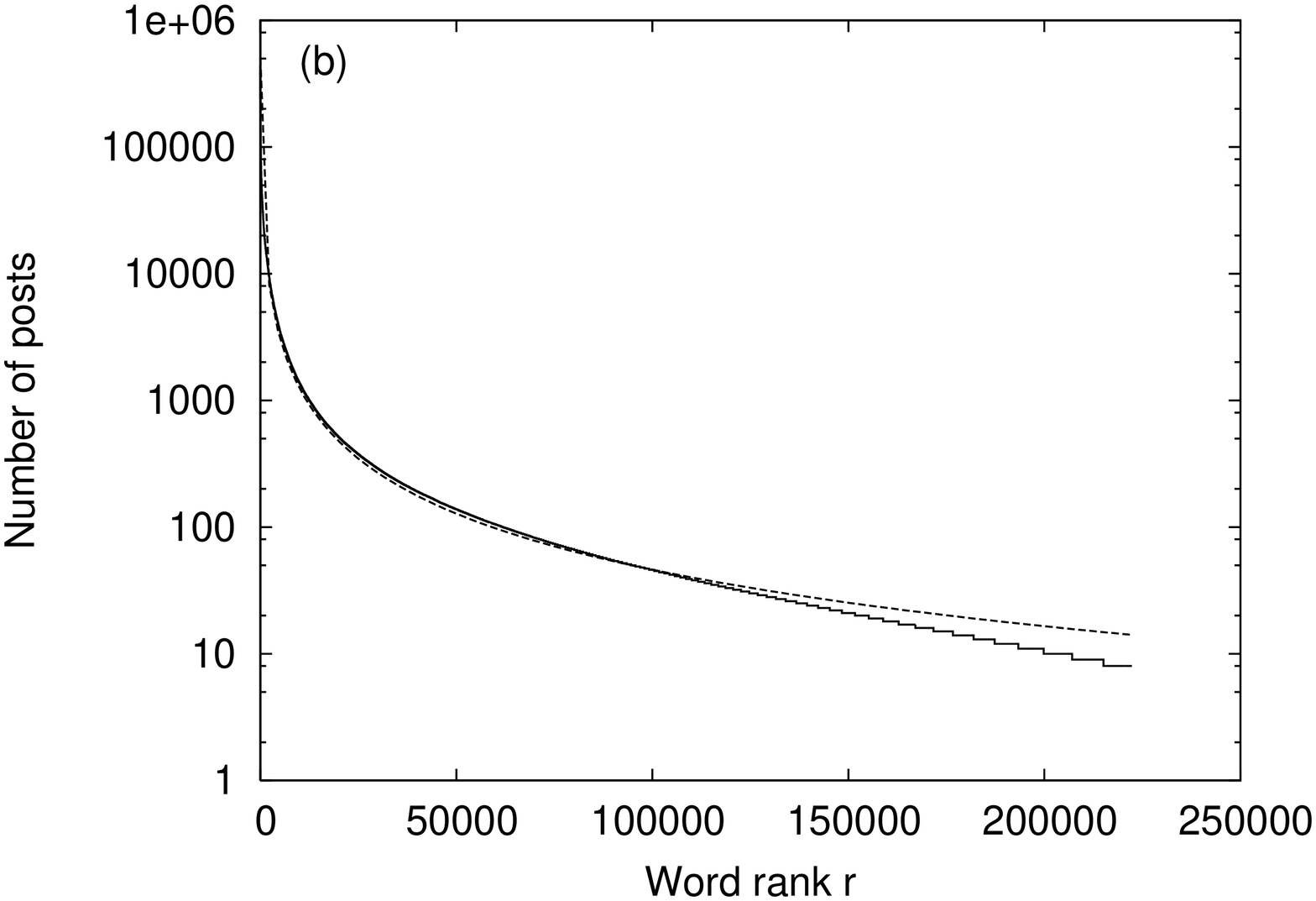}

\caption{\label{zipf}  Total number of posts containing a specific word as a function of its rank, in log-log scale (a) and log-normal scale (b). The first ten most used words are, in decreasing order, ({\em the,  a, to, of, and, in, for, is, on, it}). The curves show deviations from a power-law, but are quite well fitted (dashed line) by the modified power-law $\sim 1/(1  + 0.2   x^{0.65} + 0.0004  x^{1.5}) $.
}
\end{figure}

\subsection{Methodology}
In the following, we focus on the data collected from 68022 RSS feeds from February 11th 2005 to October 2nd 2005. The list of feeds was obtained predominantly from Google, using its filetype:rss command, in conjunction with random mid-frequency words. The purpose of this method was to gain a wide range of types of sites supporting RSS feeds. A small proportion of the feeds, about 1\%, were extracted from manual browsing of the web and the nowadays extinct {\em completeRSS.com} web site. Altogether, the feeds are predominantly composed of Blogs, but also of other sources of online information, and they are mainly in English (estimated around 70-80\%).  At this point, it is also important to stress that the boundary between major news outlets and prominent Blogs has become blurred because the top bloggers have similar readerships as major newspapers. This difficulty justifies therefore our study of
a  heterogeneous collection, that encompasses several kinds of data sources, i.e. incorporating as well personal diary-like Blogs, professional specialist Blogs and newspaper RSS feeds. Let us also stress that one drastic event took place during the period under consideration, the London Attacks of 7 July 2005.

\begin{figure}
\includegraphics[angle=-90,width=4.5in]{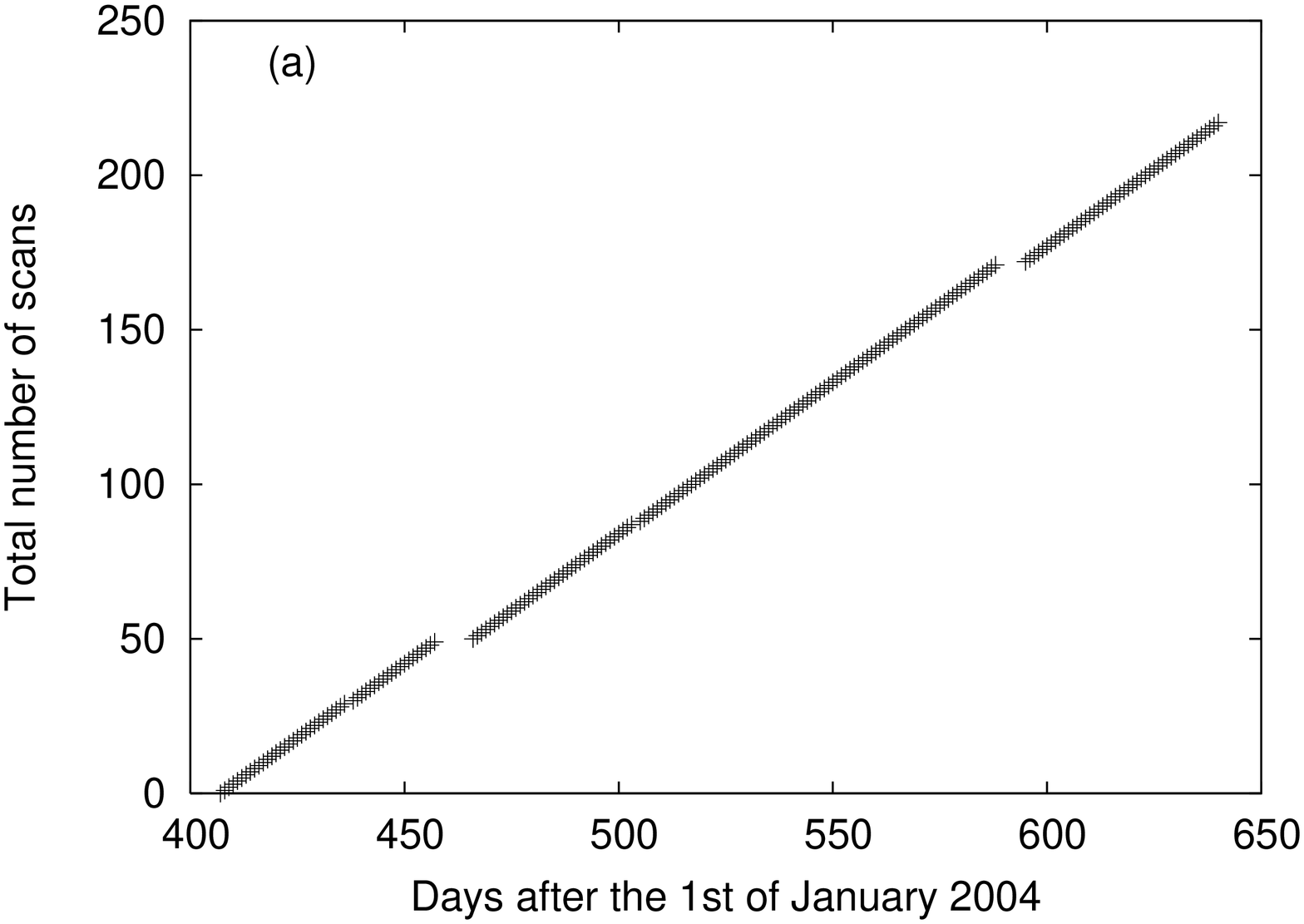}

\includegraphics[angle=-90,width=4.5in]{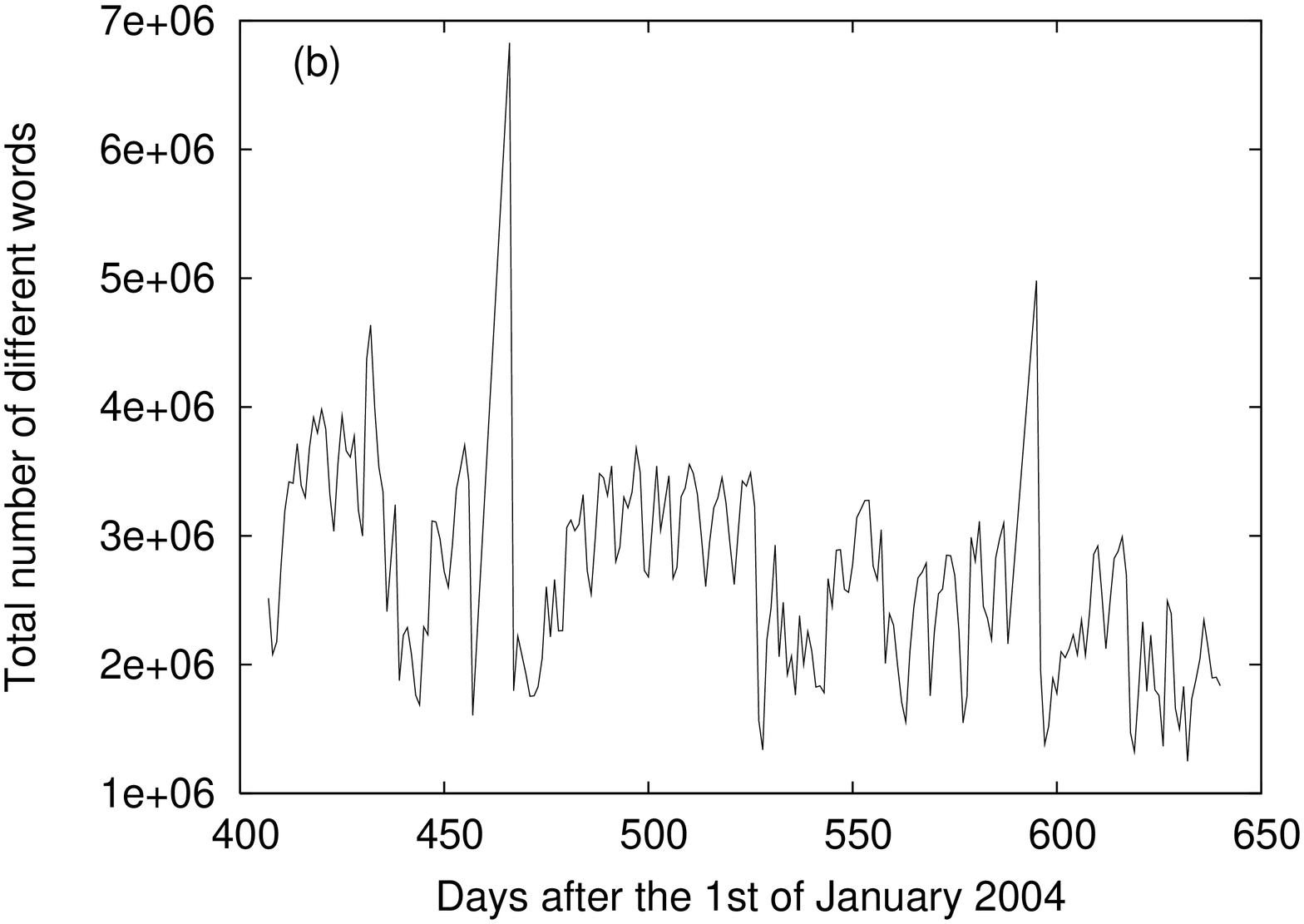}
\caption{\label{tempo}  In (a), time evolution of the number of performed scans. The discontinuities correspond to missed scans due to technical problems. In (b), we plot the measured number of different words as a function of time. Anomalous peaks are observed after each missed scan. These are removed from the analysis, as
spurious data.
}
\end{figure}

Recall that each text published by a blogger is called a post and is made of a sequence of words separated by punctuation, a blank space or  markup (e.g., HTML or XML). 
The data collection was performed as follows. Every 24 hours, all feeds were scanned and their content compared with the content observed in the last scan. All new posts are attributed to the new scanning time. Over the time period of 234 days, we observed 2294672 different words in the data set.
In Fig.1, we plot the number of posts containing a specific word as a function of its rank (the most frequently-occurring word has a rank 1, the second placed word has rank 2...). Let us remark the deviations from the power-law, i.e. Zipf law $1/x^\lambda$ and from the Zipf-Mandelbrot law $1/(1+a x)^\nu$ \cite{montem}, as those observed in \cite{deviation,raan,rousseau}. In contrast, the empirical curve of the presently examined data is very well fitted by a modified power-law of the form
\begin{equation}
\label{fita}
 1/(1  + a_1   x^{\gamma_1} + a_2  x^{\gamma_2}), 
\end{equation}
where $a_1=0.2$, $a_2=0.0004$, $\gamma_1=0.65$ and $\gamma_1=1.5$. Let us also stress that
 Eq. (\ref{fita}) includes two different characteristic exponents \cite{2exponent1,2exponent} and that it is reminiscent of Tsallis-like distributions \cite{tsallis}. The main point for the rest of this paper is that the rank function of Fig.1 behaves {\em qualitatively} like a power-law, which implies that the distribution of the number of posts also behaves like a power-law \cite{adamic}. Consequently, this distribution is very wide and not peaked around its average value, i.e. the number of posts fluctuates enormously from one word to another word.

Before going further, one should also note that the above automatic scanning has been perturbed a few times due to technical problems, leading to gaps in time as seen in Fig.2a. These missing scans have therefore led to the erroneous attribution of posts for the missing days and for the day that followed (see Fig.2b). In order to perform a time analysis of word frequencies, we removed from the time series these anomalous data. After this  data cleaning, there remained a 214 day time period. This cleaning does not change the shape of the curves of Fig.1, but reduces the systematic errors bars for the following waiting times study.

\begin{figure}
\includegraphics[angle=-90,width=4.5in]{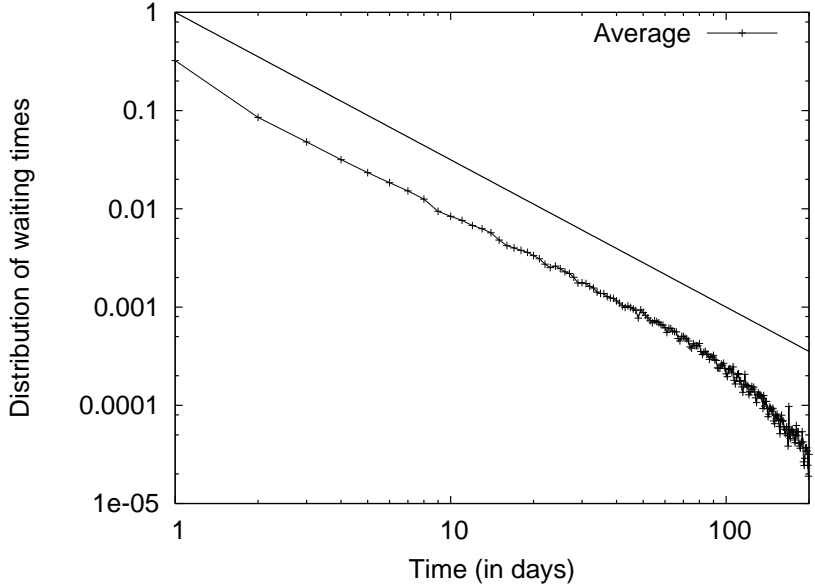}

\caption{\label{faka}  Distribution of waiting times for all words in ensembles $E_k$, $k<214$, as a function of time (in days),  in a log-log scale.  The solid line is the power-law $\tau^{-3/2}$. 
}
\end{figure}

\section{Word statistics}

\subsection{Ensembles of equivalent words}

Let us label each word by the index $\alpha$. The number of posts in which this word occurs on day $i$ is noted  $W_{\alpha i}$.  Moreover $W_\alpha = \sum_{i=1}^{214} W_{\alpha i}$ denotes the number of occurrences of $\alpha$ over the total time period. 
As discussed above, words may exhibit a large range of frequencies (1-$10^6$). The spread of these frequencies may find its origin in many causes, e.g. the word "popularity" (two synonyms may be more or less popular) or "contextuality" (words associated to general and frequent contexts should be used more often). Such effects may be estimated by typing words in Google and counting the number of matches. For instance, synonyms like "clothes" and "garments" certainly have different popularities, as their Google matches are $136 \times 10^6$ and $17 \times 10^6$ respectively. Similarly, a word associated with a popular topic/context, e.g. "music", which occurs $951 \times 10^6$ times, is used much more often than a word associated with a less popular topic, e.g. "tuberculosis" occurs $21 \times 10^6$ times.

 \begin{figure}
\includegraphics[angle=-90,width=4.5in]{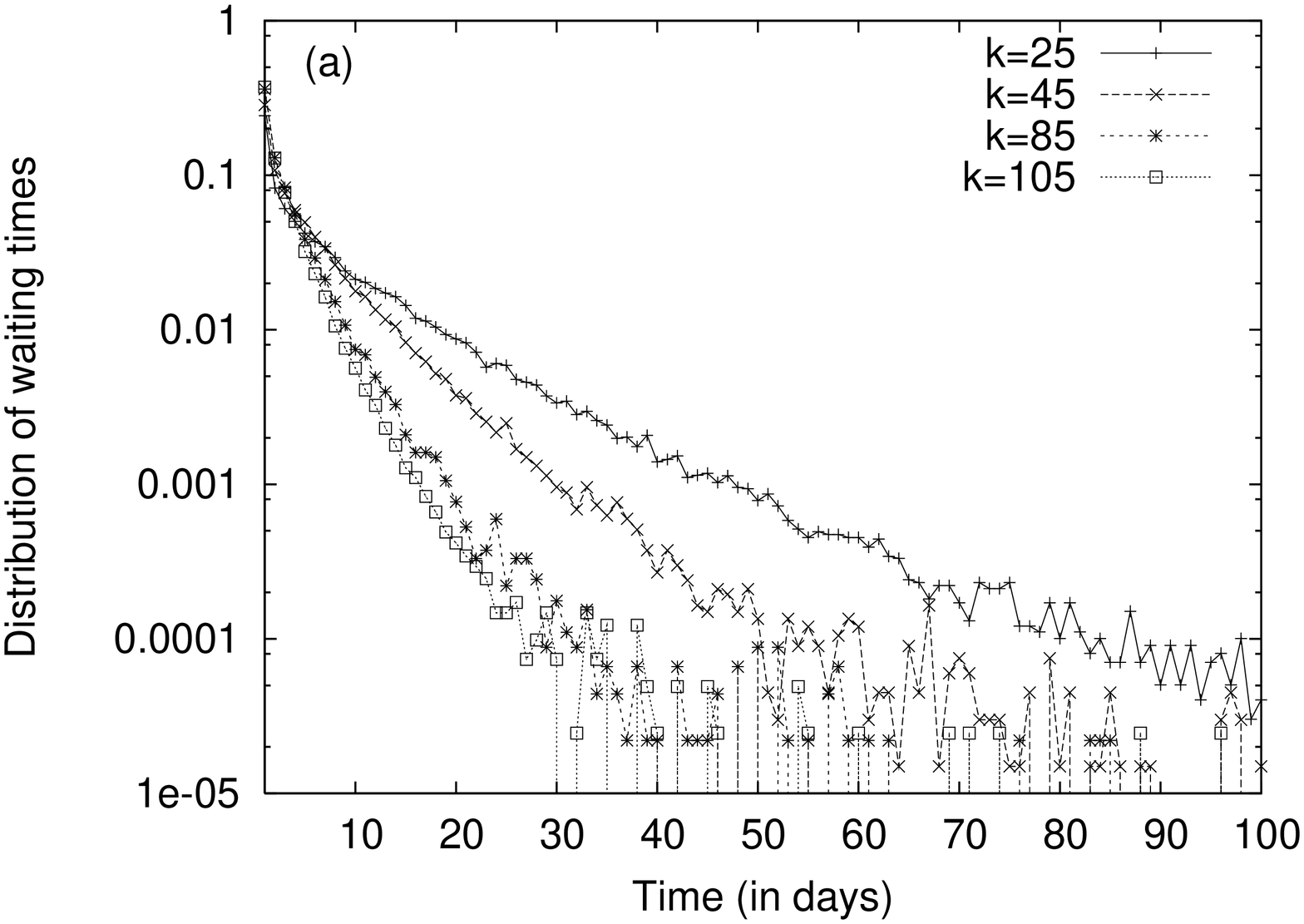}

\includegraphics[angle=-90,width=4.5in]{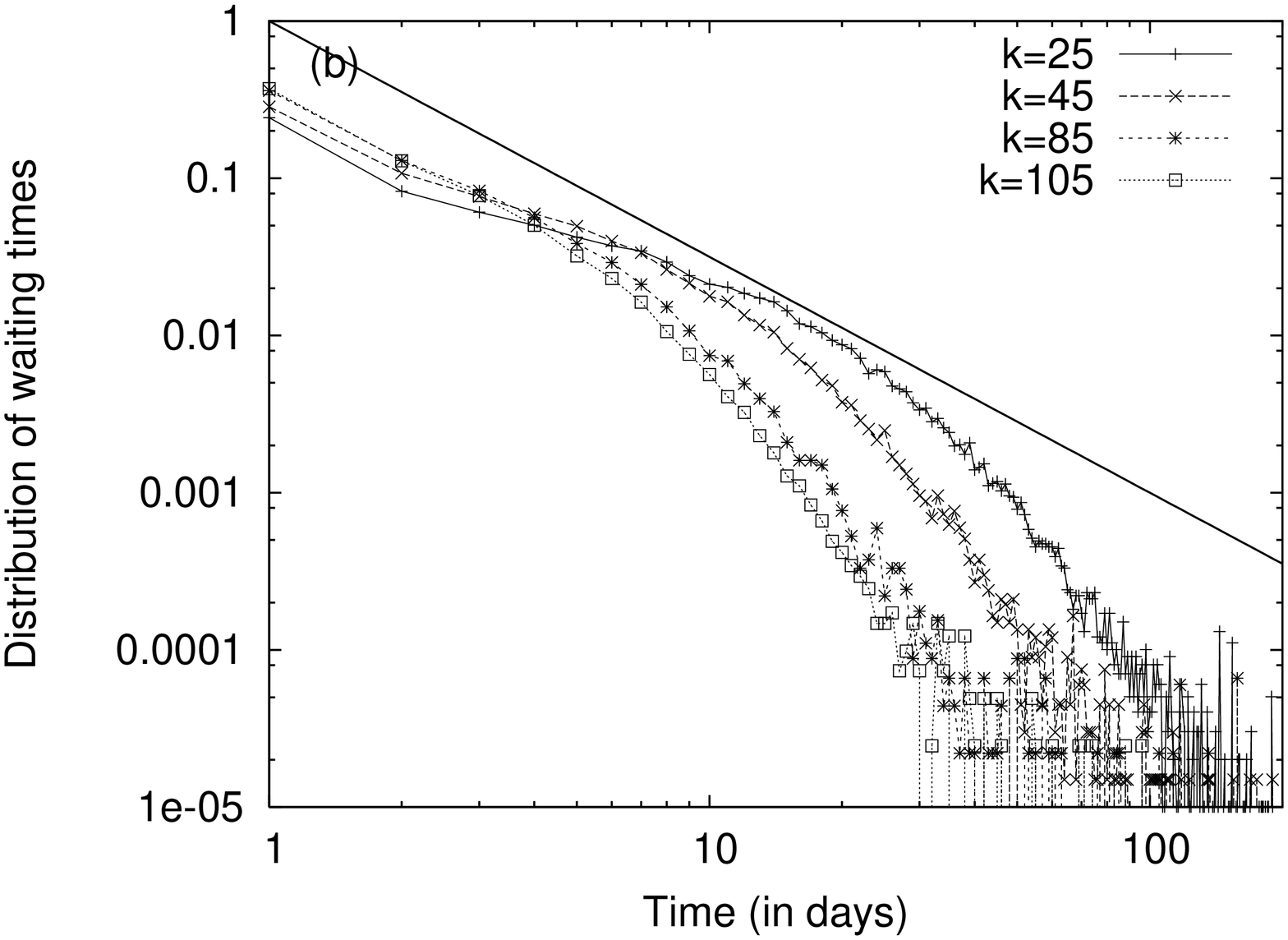}
\caption{\label{classK} Distribution of waiting times for 4 ensembles of words $E_k$, $k=[25,45,85,105]$, i.e. belonging to the dilute limit case, as a function of time (in days): (a) in a log-normal scale and (b) in a log-log scale.  The solid line is the power-law $\tau^{-3/2}$. 
}
\end{figure}

It is well-known that heterogeneous events' frequencies may artificially overpopulate the tail of the distribution of waiting times. In \cite{lambi}, for instance, it is shown that such an effect may even lead to a power-law distribution of waiting times while the system evolves in fact like a time-dependent Poisson process. This over-population originates from the fact that a distribution whose characteristic time fluctuates like
\begin{equation}
\label{fake}
f(\tau)= <\tau_{c}^{-1} \exp{(- \tau/ \tau_{c})}>_{\tau_c} \equiv \int d\tau_c \tau_{c}^{-1} \exp{(- \tau/ \tau_{c})} p(\tau_c),
\end{equation}
where $p(\tau_c)$ is the probability that the characterstic time is $\tau_c$,
 always exhibits larger fluctuations around the average waiting time than the Poisson distribution (\ref{poisson}) does \cite{super1,super2}.  
In order to overcome this difficulty, we separate out words depending on their frequencies. Define the ensemble $E_{k}$ of words $\{\alpha_{i_1}, ..., \alpha_{i_{n_k}} \}$, that occur $k$ times in the whole time interval. A word that is used only once is usually called a "hapax legomenon", while a word used twice is a "dis legomenon", thrice, a "tris legomenon", etc. Let us also denote the number of words belonging to the ensemble by $n_k$, i.e. it is the number of words $\alpha$ for which $W_\alpha=k$. In the following analysis, we consider that all words belonging to the same ensemble $E_k$ are {\em a priori} equivalent. This assumption seems reasonable {\em a priori}, as words in the same ensemble have the same average waiting time and should be more homogeneous than words randomly chosen in the whole set of used words.  The validity of our assumption will be verified {\em a posteriori} by showing that waiting times are distributed in the same way in each ensemble $E_k$.

\begin{figure}
\includegraphics[angle=-90,width=4.5in]{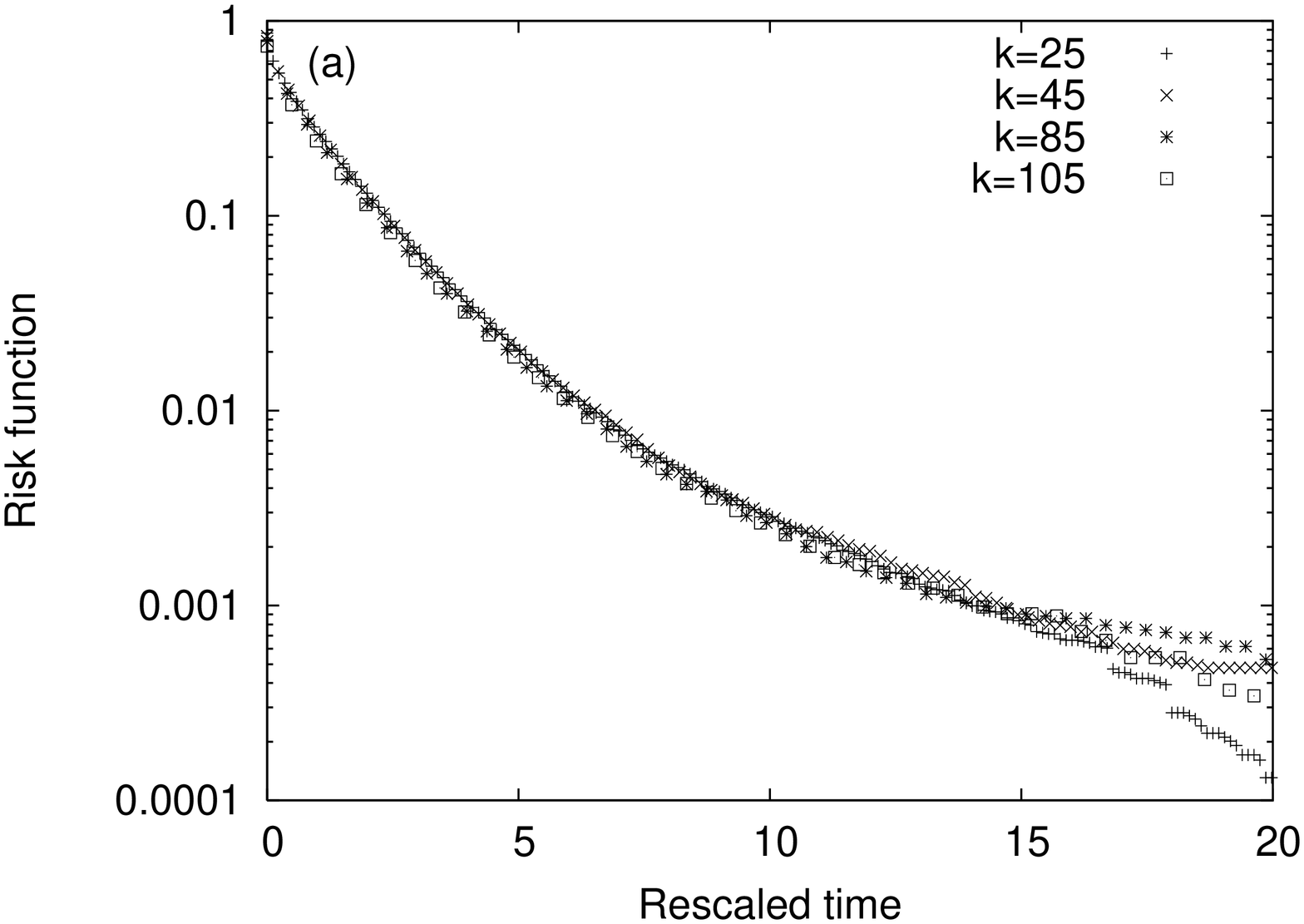}

\includegraphics[angle=-90,width=4.5in]{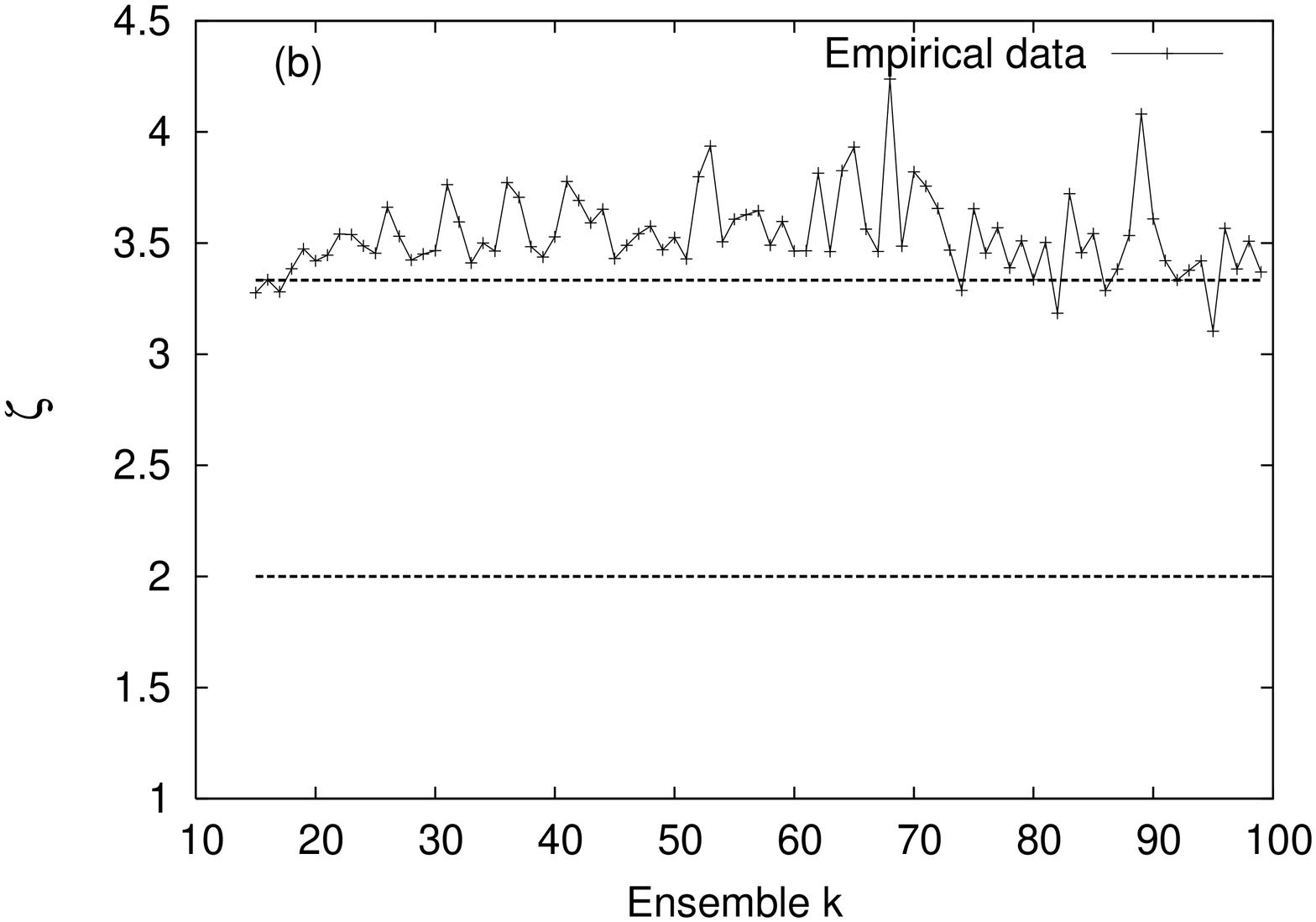}
\caption{\label{risk}  In (a), empirical Risk function for 4 ensembles of words $E_k$, $k=[25,45,85,105]$, as a function of the rescaled time $t_R$. See deviations from the exponential. In (b), we plot the quantity $\zeta$ as a function of the ensemble $k$ in which it is measured (see text for the definitions of $\zeta$ and $k$). The dashed lines point to the value for a Poisson process, i.e. $\zeta=2$, and for the stretched exponential (\ref{fitt2}), i.e. $\zeta=10/3$.
}
\end{figure}

\subsection{Dilute limit}

Very rare words, i.e. words that occur much less than once a day on average, are ideal in order to test Poisson statistics, as the exponential distribution for waiting times (2) should fit them. Consequently, we focus in this section on the ensembles of words $E_{k}$, with $k<214$, i.e. that occur on average less than once day. It is instructive to look at the distribution $f(\tau)$ (see Eq.(\ref{fake})) obtained without splitting words into classes, i.e. by averaging the distribution over all words occurring $k<214$ times. From the shape of that distribution (Fig.(\ref{faka})), one might conclude that time lags between word frequencies have a power-law distribution. We show in the following that this interpretation is erroneous and that the power-law shape is due to the averaging process described in the previous section. To do so, we measure the waiting time $\tau$ between two successive occurrences of one specific word in $E_k$, for each ensemble $E_{k}$ separately. The distribution  $f_k(\tau)$ is then obtained by performing the analysis for each word in $E_k$. It is shown (Fig.\ref{classK}) that the width of $f_k$ depends on the value of $k$ (this is expected as each ensemble $k$ is characterized by a different average frequency) and that $f_k$ produces a {\em fat tail}, i.e. anomalously large probabilities for very large and very short time intervals. This fat tail suggests that word dynamics are dominated by bursts of activities \cite{bara} followed by long periods of rest in which the word does not appear. However, contrary to the distribution of Fig.\ref{faka}, the distributions $f_k$ are not well-fitted by a power-law but resemble stretched exponentials 
\begin{equation}
\label{fitt}
f_k(\tau) = C e^{- (a \tau)^{\nu}},
\end{equation}
where $\nu$ determines the shape of the distribution, $C$ is a constant of integration and $a$ determines the time scale - which all could be dependent on $k$.
However, the exponent $\nu$ is always found to be very close to $\nu=1/2$ for all the values of $k$. In that case, the constant of integration is $C=a/2$.

Let us now show that the shape of the distributions $f_k(\tau)$ is universal. To do so, it is helpful to consider the Risk function $R_k(t)$

\begin{equation}
\label{risk}
R_k(t)=\sum_{\tau=t}^{\infty} f_k(\tau) 
\end{equation}
in order to improve the statistics. The quantity 
 $R_k(t)$ converges to zero for $t \rightarrow \infty$ in the same way the time distribution $f(\tau)$ does for the  usual  exponential and power-law time statistics \cite{lambi}.
By construction, words in the ensemble $E_k$ are used $k$ times in 214 days. Consequently, since the average waiting time $<\tau>_k$ of such words is
\begin{equation}
 <\tau>_k = \sum \tau f_k(\tau) \sim \frac{214}{k},
 \end{equation}
we change the time scale like $t \rightarrow t_R=t / (214/k)$. 
Empirical results for a large range of values of $k$ as a function of $t_R$ are shown in Fig.5a and highlight deviations from the pure exponential, thereby confirming that correlations between word occurrences do not fit the Poisson hypothesis. Moreover, one observes that curves  overlap for every $k$, thereby showing that the non-Poisson distributions are universal and that words belonging to different ensembles $E_k$ share the same statistical properties. Note that this is observed over a large range of $k\in[25,105]$. 

In order to quantify the deviations from the exponential (2), it is useful to introduce the quantity
\begin{equation}
\zeta = <\tau^2>/<\tau>^2,
\end{equation}
where the average is performed over the distribution of waiting times.
It is easily shown that $\zeta_{poisson}=2$ when the process is Poisson, while it is larger than 2 if the fluctuations around the average waiting time are larger than those of a Poisson process. If the word occurrences were periodic, this quantity would go to zero. We have measured $\zeta$ for different ensembles $E_k$, $k\in[25,105]$. It is shown in Fig.5b that the empirical value is always larger than the Poisson value 2 and that it fluctuates around $\zeta_{empirical}=3.5$. Interestingly, $\zeta_k$ does not depend on the ensemble $E_k$ in which it is measured, which implies that the fluctuations around the average waiting time are the same in all ensembles and therefore confirms the universality observed in Fig.5a. Let us stress that the empirical value $\zeta_{empirical}=3.5$ is very close the value of $\zeta$ obtained from the observed distribution (\ref{fitt}). Indeed, it is straightforward to show that $<\tau> = 6/a$ and $<\tau^2>=120/a^2$ when 
\begin{equation}
\label{fitt2}
f_k(\tau) = \frac{a}{2} e^{- (a \tau)^{1/2}},
\end{equation} 
so that $\zeta=10/3$ in that case. 
\subsection{Dense limit}

\begin{figure}
\includegraphics[angle=-90,width=5.5in]{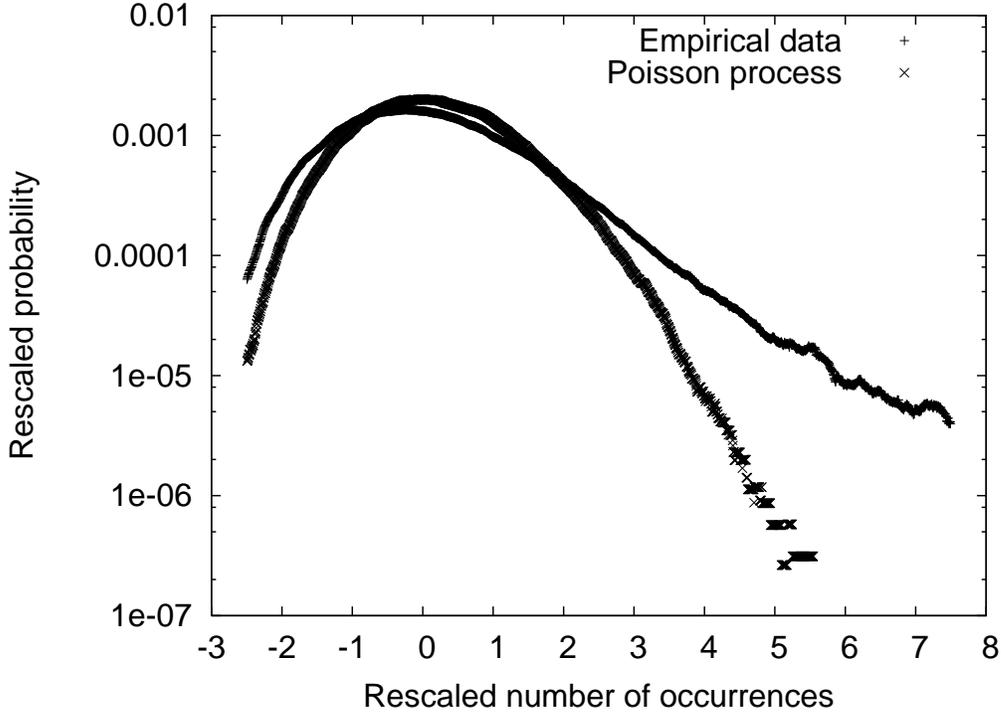}
\caption{\label{zipf} Rescaled probability distribution of the rescaled number of occurrences Eq.(\ref{oc}) (+) that measures the deviations to the average $<x>$. The data were obtained by averaging with the proper rescaling over all words occurring $k\in[1000,2000]$ times, i.e. belonging to the dense limit case.  This scheme has also been applied to Poisson random data numerically generated for the same values of $k$ ($\times$).
}
\end{figure}

For words occurring many times a day,  it is rather meaningless to focus on the time lags between their occurrences, while a statistical analysis of the number of occurrences per day makes sense. In this limit ($k \gg 1$), however, the number $n_k$ of words occurring $k$ times is very low (see Fig.1), so that a smoothing method is needed. Define $p(x,k)$ to be the probability that a word occurs $x$ times one day, if it occurs $k$ times over the $214$ days. By definition, the average number of day occurrences is $<x>=k/214$, but the width of the distribution is also expected to vary with $k$. From our data set, we  can
verify that the mean square displacement behaves like  $\sigma \sim k^{-1/2}$, as expected. These two relations suggest to focus on the rescaled variable:
\begin{equation}
\label{oc}
\tilde{x} = \frac{(x-\frac{k}{214})}{\sigma}
\end{equation}
and to the corresponding rescaled probability distribution.
By doing so, the data are smoothened and the characterization of the probability shape is possible. 

In order to compare with Poisson events, we have generated numerically random ensembles $E_k$. This was done by randomly allocating $k$ events into 214 boxes. The following step consists in measuring the distribution $p(x,k)$ and performing the above rescaling. As shown in Fig.6,  
empirical data are less peaked around the average value, i.e. extreme events happen much more often than in the Poisson case. This over-representation leads to conclusions similar to those made in the previous section. In other words, even in the dense limit,  bursts of activities occur.
 
\section{Conclusion}
 
In this article, we have performed an empirical analysis of the word frequencies arising in Blogs and RSS feeds. To do so, we have  
collected RSS data during a large time period (more than 200 days during spring 2005). These data encompass several kinds of information sources, such as newspaper RSS feeds and personal diary-like Blogs. Our analysis has been performed by discriminating words depending on their number of occurrences $k$. Namely, ensembles $E_{k}$ of words occurring with the same frequency are defined and all words belonging to that ensemble are assumed to be "equivalent". This method is especially suitable when the frequency of word occurrences is very heterogeneous during the whole time window, as an heterogeneity of frequencies may radically alter the statistics of word occurrences. 

Two limits have been considered: a dilute limit that consists of sparsely used words and a dense limit of words used many times a day. In the dilute limit, we have analyzed the statistics of waiting times between two successive occurrences of a word.  It has been shown by using a proper rescaling that the distribution is the same for many ensembles $E_k$, thereby revealing a universal behaviour for word statistics. This universal distribution of waiting times has also been shown to deviate from the pure exponential, i.e. it behaves like a stretched exponential, and a statistical quantity $\zeta$ has been introduced in order to measure these deviations.  Deviations from the Poisson distribution are also observed for the number of word occurrences per day in the dense limit. Altogether, these deviations are associated with {\em fat tails}, e.g., a high probability to observe extreme events,
and suggest that word usage is dominated by bursts of activities followed by long periods of rest. Such bursts, which have also been observed  in other social systems, e.g., Internet traffic \cite{willinger}, email or web browsing \cite{bara}, may be caused by a response to an external triggering factor (e.g., US elections, publicity) \cite{lambiShock} or arise due to active endogenous discussions between bloggers.

Theoretical models reproducing the above empirical behaviour would be of great interest. Possible interesting ingredients include aging mechanisms \cite{tags,tags2,tags3} that favour the realization of the most recent words as well as copying mechanisms in which people would have a tendency to use the words used by their acquaintances \cite{copy1,copy2,copy3,copy4}.
 
 {\bf Acknowledgements}
This work 
has been supported by European Commission Project 
CREEN FP6-2003-NEST-Path-012864.

\end{document}